\documentclass[twocolumn,showpacs,preprintnumbers,amsmath,amssymb]{revtex4}
\usepackage{epsf} 
\usepackage{graphicx}
\usepackage{dcolumn}
\usepackage{bm}

\newcommand{\bee}{\begin{equation}}
\newcommand{\ee}{\end{equation}}
\newcommand{\beea}{\begin{eqnarray}}
\newcommand{\eea}{\end{eqnarray}}

\begin{document}

\preprint{SNUTP-03-100}

\title{
  Penguin diagrams for improved staggered fermions
}
\author{Weonjong Lee}
\email{wlee@phya.snu.ac.kr}
\affiliation{
  School of Physics,
  Seoul National University,
  Seoul, 151-747, South Korea
  }
\date{\today}
\begin{abstract}
We calculate, at the one-loop level, penguin diagrams for improved
staggered fermion operators constructed using various fat links.
The main result is that diagonal mixing coefficients with penguin
operators are identical between the unimproved operators and the
improved operators using such fat links as Fat7, Fat7+Lepage,
$\overline{\rm Fat7}$, HYP (I) and HYP (II).
In addition, it turns out that the off-diagonal mixing vanishes for
those constructed using fat links of Fat7, $\overline{\rm Fat7}$ and
HYP (II).
This is a consequence of the fact that the improvement by various fat
links changes only the mixing with higher dimension operators and
off-diagonal operators.
The results of this paper, combined with those for current-current
diagrams, provide complete matching at the one-loop level with all
corrections of ${\cal O}(g^2)$ included.
\end{abstract}
\maketitle

\section{Introduction \label{sec:intr}}
Decoupling of heavy particles in the standard model leads to the
low-energy effective Hamiltonian through OPE (operator product
expansion) which achieves a factorization of short- and long-distance
contributions.
The effective Hamiltonian of our interest consists of $\Delta S = 1$
four-fermion operators which contain information at low energy and
the corresponding Wilson coefficients including the short-distance
physics at high energy.
The low energy effects of the electroweak and strong interactions can
be expressed as matrix elements of the four fermion operators between
hadron states.
Since the matrix elements involve non-perturbative QCD physics at low
energy, lattice QCD is well suited for their calculation.
The main advantage of using lattice QCD is that it provide a first
principle, non-perturbative estimate.
Different fermion discretizations such as Wilson, staggered,
domain-wall (DW) and overlap have been used to study the matrix
elements and to control their statistical and systematic errors.
In this work, we adopt staggered fermions to explore the
renormalization property of the operators relevant to calculating the
weak matrix elements.
Staggered fermions are an attractive choice for the calculation of
weak matrix elements.
They are computationally efficient and so simulations with three
dynamical flavors are already possible with relatively light quark
masses using the AsqTad staggered fermions \cite{ref:davies:1}.
They preserve enough chiral symmetry to protect operators of physical
interest from mixing with others of wrong chirality.
By construction they retain four tastes of doublers for each lattice
field, which in itself is not a problem.
Their main drawback is that the tastes symmetry is broken at non-zero
lattice spacing and is restored in the continuum limit.
At non-zero lattice spacing, quark-gluon interactions violate the
taste symmetry.
This has two consequences: first, there are large ${\cal O}(a^2)$
discretization errors in hadron spectrum and weak matrix elements and
second, some one-loop corrections are so large that matching factors
differ significantly from their tree-level value of unity.
Both problems are manifestly alleviated by improving staggered
fermions using ``fat'' links \cite{ref:lagae:1,ref:lepage:1}.
Taste symmetry violations in the pion multiplet are substantially
reduced \cite{ref:orginos:2,ref:anna:0} and one-loop corrections to
the matching factors for four-fermion operators are reduced from as
large as 100\% down to $\sim$ 10\% \cite{ref:wlee:5}.
Based upon the analysis of Ref.~\cite{ref:wlee:3,ref:wlee:0}, we
figured out that the perturbative corrections is smallest for a
particular type of fattening, ``HYP'' and ``$\overline{\rm Fat7}$''
smeared links.
Hence, we choose the ``HYP/$\overline{\rm Fat7}$'' improvement scheme
for our numerical study on the weak matrix elements \cite{ref:wlee:8}.
Recently, there have been elaborate efforts to do the higher loop
calculation for highly improved staggered fermions using the Luscher
and Weisz method \cite{ref:luscher:1}.
They automated the generation of the Feynman rules for essentially
arbitrary complicated lattice actions such as the AsqTad type
\cite{ref:trottier:1} and some even developed an algorithm which
generates Feynman diagrams automatically \cite{ref:mason:1}.
An essential step for using lattice QCD is to obtain the relationship
between the continuum and lattice four-fermion operators.
There are two classes of Feynman diagrams at the one-loop level: (1)
current-current diagrams and (2) penguin diagrams
\cite{ref:buras:0,ref:sharpe:2}.
At the one-loop level, the contribution from the current-current
diagrams and the penguin diagrams can be treated separately.
In the case of the current-current diagrams, the matching coefficients
for the operators constructed using improved staggered fermions
(HYP/$\overline{\rm Fat7}$) were presented in \cite{ref:wlee:5}.
%

%
Here, we focus on penguin diagrams in which one of the quarks in the
four-fermion operator is contracted with one of the anti-quarks to form
a closed loop.
The main goal of this paper is to calculate the penguin diagrams for
improved staggered fermion operators constructed using various fat
links and to present the matching formula between the continuum and
lattice operators at the one-loop level.
The results are compared with those for the unimproved staggered
fermions given in \cite{ref:sharpe:2}.
The role of improvement is reviewed from the standpoint of operator
mixing.
The main results of this paper, combined with those of current-current
diagrams given in Ref.~\cite{ref:wlee:5}, provide a complete set of
matching formula for $\epsilon'/\epsilon$ at the one-loop level with
all the $g^2$ corrections included.
%

%
This paper is organized as follows.
In Sec.~\ref{sec:notation+feynman-rules}, we describe our notation for
the improved staggered fermion operators and various fat links. We
also explain the Feynman rules.
In Sec.~\ref{sec:penguin}, we explicitly calculate penguin diagrams
step by step and the main results are summarized in a theorem.
We close with some conclusions in Sec.~\ref{sec:conclude}.
A preliminary result of this paper appeared in \cite{ref:wlee:7}.
%

%
%
%
\section{Notation and Feynman rules}
\label{sec:notation+feynman-rules}
Basically, the improved staggered fermion action has the same form as
for the unimproved staggered fermions \cite{ref:smit:0}, except in that
the original thin links $U_\mu$ are replaced by fat links $V_{\mu}$.
%
%
%
\begin{eqnarray}
S &=& a^4 \sum_{n} \bigg[ \frac{1}{2a}
\sum_{\mu} \eta_{\mu}(n)
\Big(
\bar{\chi}(n) V_{\mu}(n) \chi(n + \hat{\mu}) -
\nonumber \\ & & 
\bar{\chi}(n + \hat{\mu}) V^{\dagger}_{\mu}(n) \chi(n) \Big)
+ m \bar{\chi}(n)\chi(n) \bigg] \ ,
\end{eqnarray}
%
%
%
where $ n = (n_1,n_2,n_3,n_4)$ is the lattice coordinate and $
\eta_{\mu}(n) = (-1)^{n_1 + \cdots + n_{\mu-1}} $.
Here, the fat link $V_\mu$ represents collectively Fat7
\cite{ref:orginos:1}, Fat7+Lepage \cite{ref:lepage:0}, HYP links
\cite{ref:anna:0} and SU(3) projected Fat7 ($\overline{\rm Fat7}$) links
\cite{ref:wlee:0}.
The detailed definitions of various fat links are given in the
original references and so we do not repeat them here.
In addition, MILC and HPQCD collaboration developed and have used the
AsqTad staggered action
\cite{ref:orginos:1,ref:orginos:2,ref:lagae:1,ref:lepage:1}, whose
gauge action is a one-loop Symanzik improved action and whose fermion
action contains a Naik term in order to remove the $O(a^2)$
discretization errors.
Out of various choices, HYP and $\overline{\rm Fat7}$ suppress, in
particular, the taste changing gluon interactions efficiently and
reduce the taste symmetry breaking in the pion spectrum significantly
\cite{ref:anna:0}, compared with others.
In addition, it turns out that the one-loop corrections are smallest
for HYP and $\overline{\rm Fat7}$ compared with others
\cite{ref:wlee:3} and that they possess several nice properties in
renormalization which are explained in \cite{ref:wlee:0}.
If the goal of the improvement were to minimize the ${\cal O}(a^2)$
discretization error and to achieve a better scaling behavior through
the Symanzik improvement program, it would be natural to choose the
action of the AsqTad type and improve the operators correspondingly.
However, our improvement goal is to minimize the perturbative
correction as much as possible (if possible, down to less than 10\% at
the one loop level), which, in fact, turned out to be the same as to
minimize the taste symmetry breaking effect \cite{ref:golterman:1}.
Hence, we have chosen HYP/$\overline{\rm Fat7}$ improvement scheme for
our numerical study on $\epsilon'/\epsilon$ mainly because it serves
better to our purpose of improvement.
In this paper, we adopt the same notation for fat links as in
\cite{ref:wlee:0,ref:wlee:3,ref:wlee:5}.
%

In order to construct the four spin component Dirac field, we adopt
the coordinate space method suggested in
\cite{ref:klubergstern:0}.
In this method, we interpret 16 staggered fermion fields ($\chi$) of
each hypercube as 4 Dirac spin and 4 flavor (=taste) components.
The continuum limit of the staggered fermion action on the lattice
corresponds to QCD with four degenerate flavors ($ N_f = 4 $)
\cite{ref:smit:0}.
There are numerous choices to transcribe the lattice operator for a
given continuum operator \cite{ref:wlee:2,ref:sharpe:0,ref:sharpe:1}.
We adopt the same convention and notation as in \cite{ref:wlee:1}
except for the gauge links.
We denote the gauge invariant bilinear operators as
\begin{eqnarray}
[ S \times F ]
&\equiv& \frac{1}{N_f} \sum_{A,B}
[\bar{\chi_a}(y_A) \ 
(\overline{ \gamma_S \otimes \xi_F } )_{AB} \ 
\chi_b(y_B)] 
\nonumber \\ & & 
\  {\cal V}^{ab}(y_A,y_B)
\end{eqnarray}
where $ S $ and $ F $ represent spin and flavor (=taste) respectively
and correspond to one of the following; $ S $ (scalar), $ V_\mu $
(vector), $ T_{\mu\nu} $ (tensor), $ A_\mu $ (axial) and $ P $
(pseudo-scalar).
Here, $ {\cal V}(y_A,y_B) $ is a product of gauge links that makes the
bilinear operator gauge invariant.
The link matrices $ {\cal V}(y_A,y_B) $ are constructed by averaging
over all of the shortest paths between $ y_A $ and $ y_B $, such that
the operator $[ \gamma_S \times \xi_F ]$ is as symmetric as possible:
\begin{eqnarray}
\label{ulink}
{\cal V}(y_A,y_B) & = & \frac{1}{4!}
\sum_{P} V(y_A, y_A+\Delta_{P1})\cdots
\nonumber \\ & &
V(y_A+\Delta_{P1}+\Delta_{P2}+\Delta_{P3}\ ,\ y_B) \ ,
\end{eqnarray}
where
$ P $ is an element of the permutation group (1234) and
\begin{eqnarray}
\label{delta}
\Delta_{\mu} = (B_{\mu}-A_{\mu}) \hat{\mu} \ .
\end{eqnarray}
For the four-fermion operators, we use the same notation as the
bilinears but need to distinguish between color one trace and color
two trace operators.
%
\begin{eqnarray}
[ S \times F ] [ S' \times F' ]_{I} 
&\equiv& 
 \frac{1}{N_f^2} \sum_{A,B,C,D}
\nonumber \\ & & 
[\bar{\chi}(y_A) \  
(\overline{ \gamma_S \otimes \xi_F } )_{AB} \ 
\chi(y_B) ] 
\nonumber \\ & &
\cdot 
[\bar{\chi}(y_C) \ 
(\overline{ \gamma_{S'} \otimes \xi_{F'} } )_{CD} \ 
\chi(y_D) ] 
\nonumber \\ & &
\cdot {\cal V}(y_A,y_D) \ {\cal V}(y_C,y_B)
\\ {}
[ S \times F ] [ S' \times F' ]_{II} 
&\equiv& 
\frac{1}{N_f^2} \sum_{A,B,C,D}
\nonumber \\ & &
[\bar{\chi}(y_A) \  
(\overline{ \gamma_S \otimes \xi_F } )_{AB} \ 
\chi(y_B)] 
\nonumber \\ & & \cdot
[\bar{\chi}(y_C) \ 
(\overline{ \gamma_{S'} \otimes \xi_{F'} } )_{CD} \ 
\chi(y_D)] 
\nonumber \\ & &
\cdot {\cal V}(y_A,y_B) \  {\cal V}(y_C,y_D)
\end{eqnarray}
%
Here, note that the sub-indices $I$, $II$ represent the color one
trace and color two trace operators respectively.
%
%
%
There are two completely independent methods to construct operators on
the lattice using a Fierz transformation: one spin trace formalism and
two spin trace formalism \cite{ref:wlee:2}.
In this paper, we choose two spin trace formalism to construct the
lattice operators and it is also adopted for our numerical study on
$\epsilon'/\epsilon$.
The Feynman rules for the unimproved staggered fermions are standard
and presented in \cite{ref:wlee:1,ref:wlee:2,ref:sharpe:2,ref:sharpe:3,ref:jlqcd:0,ref:sheard:0,ref:sheard:1}.
Here, we use the same notation for Feynman rules as in 
\cite{ref:wlee:1,ref:wlee:2} and we do not repeat them.
By introducing fat links, we need to change the Feynman rules.
These changes in the Feynman rules originating from fat links are
given in \cite{ref:wlee:3,ref:wlee:5}.
Here we adopt the same notation and Feynman rules as in
\cite{ref:wlee:3,ref:wlee:5}.
We explain only the essential ingredients for a one-loop calculation.
We define the gauge fields of the thin link and fat links
as
\begin{eqnarray}
& & 
U_\mu(x) = \exp 
\big( i a A_\mu ( x + \frac{1}{2} \hat{\mu} ) \big) 
\label{eq:thin:gauge}
\\ & &
V_\mu(x) = \exp 
\big( i a B_\mu ( x + \frac{1}{2} \hat{\mu} ) \big)
\label{eq:fat:gauge} 
\end{eqnarray}
Here, note that $V_\mu$ corresponds to various fat links.
We call $B_\mu$ a ``fat gauge field'' and $A_\mu$ a ``thin gauge
field''.
This fat gauge field can be expressed in terms of thin gauge
fields as follows:
\begin{eqnarray*}
B_\mu &=& \sum_{n=1}^{\infty} B_\mu^{(n)}
\\
&=& B^{(1)}_\mu + B^{(2)}_\mu + {\cal O}(A^3)
\end{eqnarray*}
Here, $B^{(n)}_\mu$ represents a term of order $A^n$.
Theorems 1 and 2 of \cite{ref:wlee:0} say that the linear term is
invariant under SU(3) projection and that since the quadratic term is
anti-symmetric in thin gauge fields, its contribution vanishes at the
one-loop level \footnote{ We heard that this is known and used in
\cite{ref:bernard:0,ref:sharpe:3}.}.
Hence, at one loop level, the renormalization of the gauge invariant
staggered composite operators constructed using fat links of HYP type
can be done by simply replacing the propagator of $A_\mu$ by that of
$B^{(1)}_\mu$.
This simplicity is extensively used to calculate the one-loop
correction to the improved staggered operators
\cite{ref:wlee:3,ref:wlee:5}.

The linear term $B^{(1)}$ can be expressed in momentum space as
\begin{eqnarray*}
B^{(1)}_\mu (k) &=&  \sum_\nu h_{\mu\nu}(k) A_\nu(k) \ .
\end{eqnarray*}
The details of the blocking transformation for the fat links are
contained in $h_{\mu\nu}(k)$, which is given in
\cite{ref:wlee:3,ref:wlee:5} for various fat links.
\begin{eqnarray}
h_{\mu\nu}(k) &=& \delta_{\mu\nu} D_\mu(k) +
(1 - \delta_{\mu\nu}) G_{\mu\nu}(k)
\nonumber \\
D_\mu(k) &=&  1 - d_1 \sum_{\nu\ne\mu} {\bar s}_\nu^2
+ d_2 \sum_{\nu < \rho \atop \nu,\rho\ne\mu}{\bar s}_\nu^2 {\bar s}_\rho^2
\nonumber \\ & &
- d_3 {\bar s}_\nu^2 {\bar s}_\rho^2 {\bar s}_\sigma^2
- d_4 \sum_{\nu\ne\mu} {\bar s}_\nu^4
\nonumber \\
G_{\mu\nu}(k) &=&
{\bar s}_\mu {\bar s}_\nu \widetilde G_{\nu,\mu}(k) \\
\widetilde G_{\nu,\mu}(k) &=& d_1
- d_2 \frac{({\bar s}_\rho^2+ {\bar s}_\sigma^2)}{2}
+ d_3 \frac{{\bar s}_\rho^2 {\bar s}_\sigma^2}{3}
+ d_4 {\bar s}_\nu^2
\label{eq:h_mat}
\end{eqnarray}
Here, the coefficients $d_i$ distinguish the different choices
of fat links.
\begin{enumerate}
\item Unimproved (naive):
\begin{equation}
d_1 = 0, \quad
d_2 = 0, \quad
d_3 = 0, \quad
d_4 = 0.
\end{equation}
\item
Fat7 links:
\begin{equation}
d_1 = 1, \quad
d_2 = 1, \quad
d_3 = 1, \quad
d_4 = 0.
\end{equation}
\item
HYP (I) links:
\begin{eqnarray}
d_1 &=& (2/3)\alpha_1(1+\alpha_2(1+\alpha_3)),
\nonumber \\
d_2 &=& (4/3)\alpha_1\alpha_2(1+2\alpha_3),
\nonumber \\
d_3 &=& 8 \alpha_1\alpha_2\alpha_3,
\nonumber \\
d_4 &=& 0.
\end{eqnarray}
We consider two choices for the $\alpha_i$. The first was determined
in \cite{ref:anna:0} using a non-perturbative optimization
procedure: $\alpha_1=0.75$, $\alpha_2=0.6$ $\alpha_3=0.3$
(we call this choice ``HYP (I)'').
This gives
\begin{equation}
d_1 = 0.89\,, \quad
d_2 = 0.96\,, \quad
d_3 = 1.08\,, \quad
d_4 = 0\,.
\end{equation}
\item
HYP (II) links:
The second is chosen so as to remove $O(a^2)$ flavor-symmetry breaking
couplings at tree level.  This choice, $\alpha_1=7/8$, $\alpha_2=4/7$
and $\alpha_3=1/4$ (we call this choice ``HYP (II)''), gives
\begin{equation}
d_1 = 1, \quad
d_2 = 1, \quad
d_3 = 1, \quad
d_4 = 0,
\end{equation}
i.e. the same as for Fat7 links.
\item
Fat7+Lepage ($O(a^2)$ improved links):
\begin{equation}
d_1 = 0, \quad
d_2 = 1, \quad
d_3 = 1, \quad
d_4 = 1,
\end{equation}
\end{enumerate}
For later convenience, we name the SU(3) projected Fat7 scheme
``$\overline{\rm Fat7}$''.
In \cite{ref:wlee:0}, we studied various fat links from the standpoint
of renormalization to improve staggered fermions and it turns out that
at the one-loop level, the renormalization of staggered fermion operators
is identical between $\overline{\rm Fat7}$ and HYP (II).
In addition, it was explained that SU(3) projection plays a role in
tadpole improvement for the staggered fermion doublers.

\section{Penguin diagrams}
\label{sec:penguin}
%
%
Here, we study penguin diagrams in which one of the quarks in the
four-fermion operator is contracted with one of the anti-quarks to form
a closed loop.
The main goal is to calculate the penguin diagrams for
improved staggered fermion operators and to provide a matching
formula between the continuum and lattice operators at the one-loop
level.
On the lattice, the gauge non-invariant four-fermion operators such as
Landau gauge operators mix with lower dimension operators which are
gauge non-invariant \cite{ref:sharpe:2}.
It is required to subtract these contributions non-perturbatively.
However, it is significantly harder to extract the divergent mixing
coefficients in a completely non-perturbative way.
This make it impractical to use gauge non-invariant operators for
the numerical study of the CP violation.
Hence, it is necessary to use gauge-invariant operators in order to
avoid unwanted mixing with lower dimension operators.
For this reason, we choose gauge-invariant operators in our numerical
study.
%

%
%
\begin{figure}
\epsfxsize=0.8\hsize\epsfbox{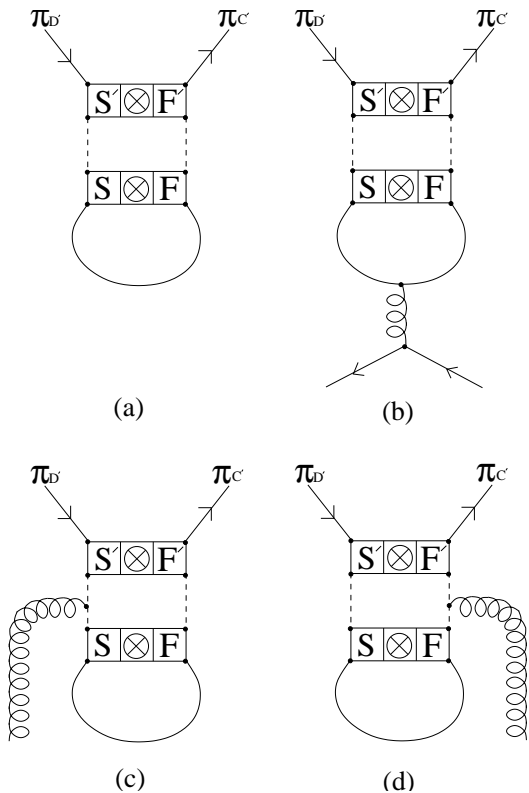}
\caption{\label{fig:1} Penguin diagrams for the staggered fermions}
\end{figure}
\begin{figure}
\epsfysize=1.05\hsize\epsfbox{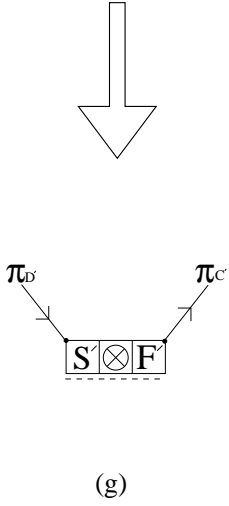}
\caption{\label{fig:3} Bilinear mixing.}
\end{figure}
\begin{figure}[t!]
\epsfxsize=0.8\hsize\epsfbox{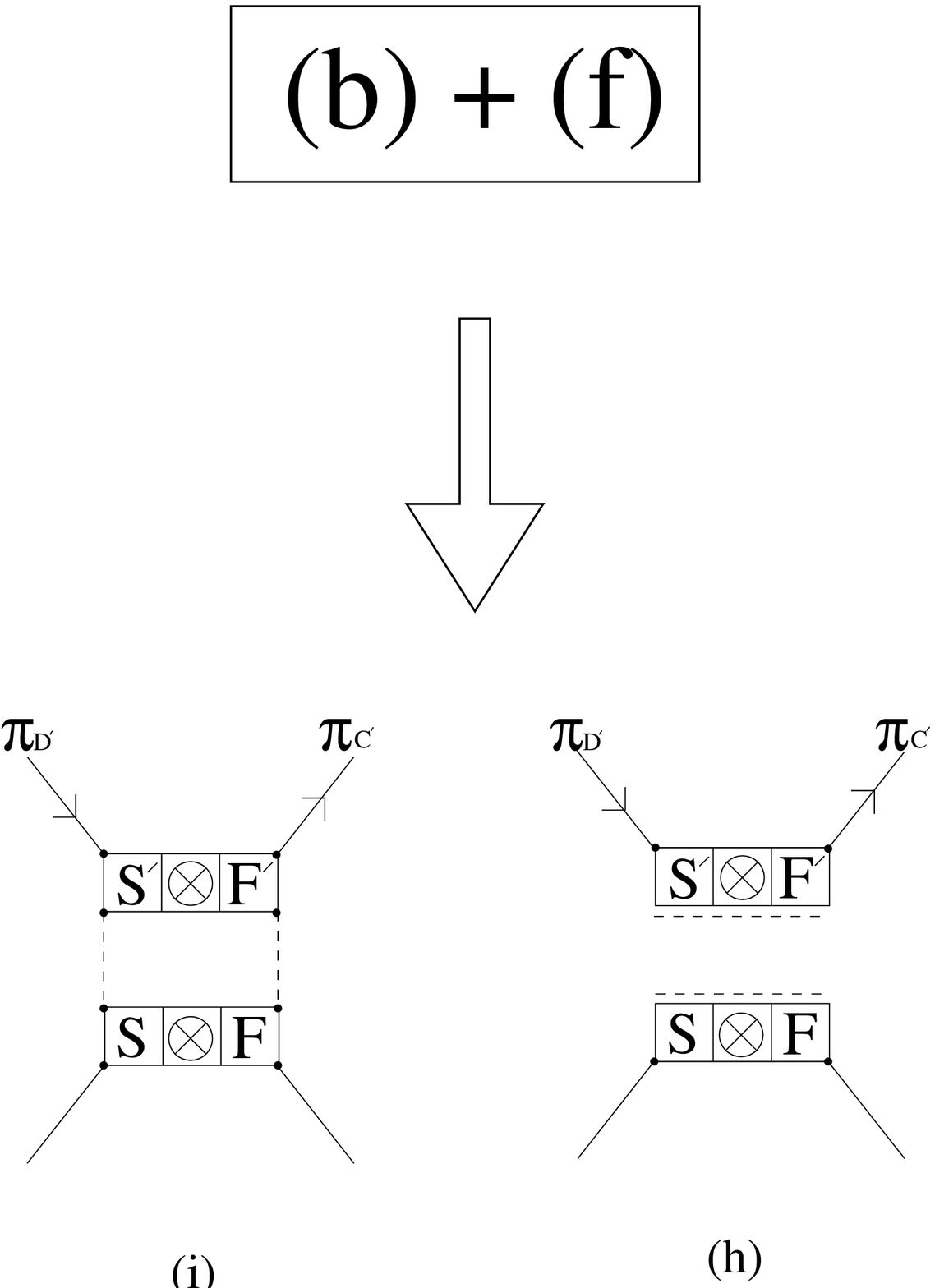}
\caption{\label{fig:4} Mixing with penguin operators}
\end{figure}

In the staggered fermion formalism there are four penguin diagrams at
the one-loop level as shown in Fig.~\ref{fig:1}.
These diagrams of penguin type can mix with lower dimension operators 
in addition to four-fermion operators of the same dimension or higher.
The mixing coefficients with lower dimension operators are
proportional to inverse powers of the lattice spacing.
The perturbation, however, is not reliable with divergent
coefficients.
Hence, we must use a non-perturbative method to determine them and
subtract away the lower dimension operators.
In the case of mixing with operators of the same dimension, the
perturbative calculation is expected to be reliable as long as the
size of the one-loop correction is small enough, which can be achieved
naturally by improving the staggered operators using fat links.
In Fig.~\ref{fig:1}, diagrams (a) and (b) have their correspondence in
the continuum and diagrams (c) and (d) do not have any continuum
correspondence.
However, diagrams (c) and (d) play an essential role in keeping
the gauge invariance in the final sum.
In other words, the gauge invariance is broken without them. 
\begin{figure}
\epsfxsize=0.8\hsize\epsfbox{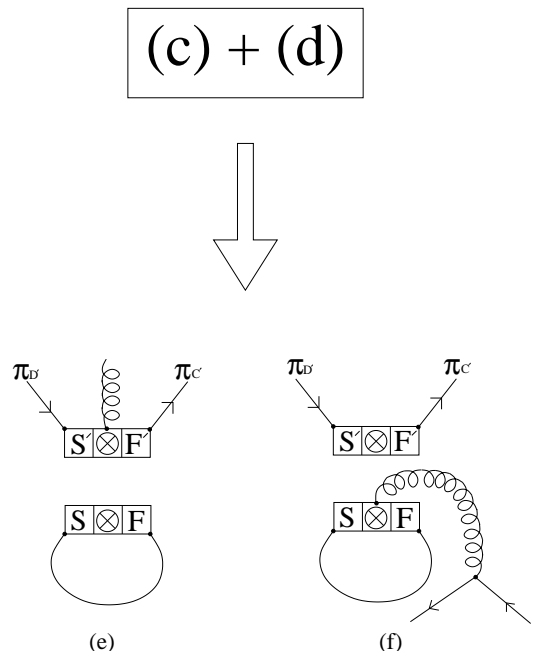}
\caption{\label{fig:2} Diagram identity: (c)+(d) = (e)+(f).}
\end{figure}
First, we overview the role of each diagram in the gauge invariance
and present the details later.
Basically, the contribution from diagrams (c) and (d) can be
re-expressed as a sum of two separate terms: diagram (e) and (f) in
Fig.~\ref{fig:2}.
We observe that the sum of diagram (a) of Fig.~\ref{fig:1} and diagram
(e) of Fig.~\ref{fig:2} produces bilinear operators in a gauge
invariant form as shown in Fig.~\ref{fig:3}.
It turns out that the contribution from diagram (b) in
Fig.~\ref{fig:1} and diagram (f) in Fig.~\ref{fig:2} leads to
four-fermion operators of our interest in a gauge invariant form,
which are typically called ``Penguin diagrams'' in the literature
as shown in Fig.~\ref{fig:4}.
Once more, we emphasize that diagrams (c) and (d) (or equivalently (e)
and (f)) are essential to keep the gauge invariance.

\subsection{Penguin operators on the lattice}
In order to construct a lattice version of continuum penguin
operators, we need some guide lines, because the staggered fermions
carry four degenerate tastes by construction, unlike the continuum
fermions.
Hence, a closed loop of staggered quarks contains four degenerate
tastes running around it, rather than a single quark, which needs to
be normalized properly by $1/N_f = 1/4$.
Penguin diagrams occur for operators which belongs to octet irrep of
the continuum flavor SU(3) symmetry.
Apart from the overall factor, the calculation is identical for all
the penguin operators of our interest \cite{ref:buras:1,ref:sharpe:2}.
Therefore, we may choose the following operator as a representative
without loss of generality:
\begin{equation}
{\cal O}^{\rm Cont}_{S',S}
\quad \longleftrightarrow \quad
{\cal O}^{\rm Latt}_{S'F',SF}
\end{equation}
where the operators are defined as 
\begin{eqnarray*}
{\cal O}^{\rm Cont}_{S',S} &=& 
[ \bar{\psi}_s \gamma_{S'} \psi_d ] 
[ \bar{\psi}_u \gamma_{S} \psi_u ]
\nonumber \\
{\cal O}^{\rm Latt}_{S'F',SF} &=& \frac{1}{N_f^2} 
[ \bar{\chi}_s \overline{ ( \gamma_{S'} \otimes \xi_{F'} ) } \chi_d ] 
[ \bar{\chi}_u \overline{ ( \gamma_{S} \otimes \xi_{F} ) } \chi_u ]
\end{eqnarray*}
Here, we adopt the two spin trace formalism \cite{ref:wlee:2}, color
contractions with gauge links are dropped for brevity, and the
subscripts $s,d,u$ represent the continuum quark flavors.
Here we select the continuum flavors so that there is only one
possibility of up quark contraction to form a closed loop, leading to a
penguin diagram.
The bilinear with strange and down quarks behaves as a spectator in
the calculation.
This choice of continuum flavor assignment was suggested originally in
\cite{ref:sharpe:2}.

Penguin diagrams of the above operators lead to mixing with the same
class of SU(3) octet operators (penguin operators):
\begin{eqnarray}
{\cal O}^{\rm Cont,P}_{S',S} &=& 
[ \bar{\psi}_s \gamma_{S'} \psi_d ] 
\sum_q [ \bar{\psi}_q \gamma_{S} \psi_q ]
\\
{\cal O}^{\rm Latt,P}_{S'F',SF} &=& 
\nonumber \\ & & \hspace*{-10mm}
\frac{1}{N_f^2} 
[ \bar{\chi}_s \overline{ ( \gamma_{S'} \otimes \xi_{F'} ) } \chi_d ] 
\sum_q [ \bar{\chi}_q \overline{ ( \gamma_{S} \otimes \xi_{F} ) } \chi_q ]
\end{eqnarray}
where the sums run over the active light flavors such as $u,d,s$.

We adopt the same notation as in \cite{ref:wlee:1,ref:sharpe:2}
to incorporate two different color contractions:
\begin{equation}
\vec{\cal O} = \left( \begin{array}{c} 
  {\cal O}_I \\ {\cal O}_{II} 
\end{array} \right)
\end{equation}
The contribution from the color two trace operator, ${\cal O}_{II}$
vanishes in the penguin diagram because it is proportional to ${\rm
Tr}(T^a) = 0$ ($T^a$ is the SU(3) group generator which is traceless).
Hence, it is sufficient to work only on the color one trace operator,
${\cal O}_{I}$.
This simplifies the matching formula.
\begin{equation}
({\cal O}^{\rm Cont}_i)_I = ({\cal O}^{\rm Latt}_i)_I 
+ \frac{g^2}{(4\pi)^2} \sum_j Z_{ij} 
\Big( \vec{P} \cdot \vec{\cal O}^{\rm Latt}_j \Big)
\end{equation}
where $i,j$ includes both the spin and taste indices, and
the projection vector is
\begin{equation}
\vec{P} = ( + \frac{1}{2}, - \frac{1}{6} )
\end{equation}
The $Z_{ij}$ have two separate contributions: one from the continuum
operators and the other from the lattice operators.
\begin{equation}
Z_{ij} = Z^{\rm Cont}_{ij} - Z^{\rm Latt}_{ij}
\end{equation}
The main goal of this paper is to calculate $Z^{\rm Latt}_{ij}$,
since $Z^{\rm Cont}_{ij}$ is well known \cite{ref:bernard:1}.
\subsection{Feynman diagrams}
\label{sec:diagram}
Using the Feynman rules described in
Sec.~\ref{sec:notation+feynman-rules}, we express the analytical form
of the Feynman diagrams presented in Fig.~\ref{fig:1}.
%
%
Diagram (a) of Fig.~\ref{fig:1} contains only mixing with bilinears
and combined with diagram (e) of Fig.~\ref{fig:2} makes the bilinears
gauge invariant as shown in Fig.~\ref{fig:3}.
\begin{widetext}
\begin{eqnarray}
G_{(a)} &=& - \frac{1}{N_f} H^{I(a)}_\mu
\\
H^{I(a)}_\mu &=&
\delta_{cd} \ 
\overline{ \overline{ (\gamma_{S'} \otimes \xi_{F'}) } }_{C'D'}
\ \frac{1}{16} \sum_{AB} 
\overline{ ( \gamma_S \otimes \xi_F ) }_{AB} 
\cdot I^{(a)}_{AB}
\\
I^{(a)}_{AB} &=& \int_p \ \exp( i p \cdot (A-B) ) \ [ S_F(p)]_{BA}
\label{eq:I(a):1}
\end{eqnarray}
\end{widetext}
Diagram (b) of Fig.~\ref{fig:1}, combined with diagram (f), leads to
mixing with four-fermion operators as shown in Fig.~\ref{fig:4}.
Using the Feynman rules described in Sec.~\ref{sec:notation+feynman-rules}
we can express the diagram (b) as follows:
\begin{widetext}
\begin{eqnarray}
G_{(b)} &=& - \frac{1}{N_f} \ \int_k \  H^{I,(b)}_\mu(k) \cdot
D^{IJ}_{\mu\nu}(k) \cdot 
V^J_\nu (p + \pi_C, -q-\pi_D, -k)
\\
H^{I,(b)}_{\mu} (k) &=&
( - i g ) \ T^I_{cd} \ 
\overline{ \overline{ (\gamma_{S'} \otimes \xi_{F'}) } }_{C'D'}
\frac{1}{16} \sum_{AB} \overline{ ( \gamma_S \otimes \xi_F ) }_{AB}
\exp\Big( i \frac{k}{2} \cdot (A+B) \Big) 
\cdot I^{(b)}_{AB} (k)
\\
I^{(b)}_{AB}(k) &=& 
\int_p \ \exp\Big( i p \cdot (A-B) \Big) \ 
\Big\{ [ S_F(p-\frac{k}{2}) ] 
\cdot \cos(p_\mu) \overline{ ( \gamma_\mu \otimes 1 ) }
\cdot [ S_F(p+\frac{k}{2}) ] \Big\}_{BA}  
\end{eqnarray}
\end{widetext}
%
%
Diagram (c) and (d) of Fig.~\ref{fig:1} contains mixing with bilinears
and four-fermion operators as shown in Fig.~\ref{fig:2}.
These two diagrams play an essential role in keeping the gauge
invariance of the final results.
\begin{widetext}
\begin{eqnarray}
G_{(c)} &=& - \frac{1}{N_f} H^{I,(c)}_\mu (k)
\\
H^{I,(c)}_\mu (k) &=&  (+ig) \ T^I_{cd} 
\frac{1}{16} \sum_{A,B} \overline{ (\gamma_S \otimes \xi_F) }_{AB}
\frac{1}{16} \sum_{A',B'} (-1)^{C \cdot A'}
\overline{ (\gamma_{S'} \otimes \xi_{F'}) }_{A'B'} (-1)^{D \cdot B'}
(B-A')_\mu f^\mu_{A'B} (k) 
\ I^{(a)}_{AB}
\\
G_{(d)} &=& - \frac{1}{N_f} H^{I,(d)}_\mu (k)
\\
H^{I,(d)}_\mu (k) &=&  (+ig) \ T^I_{cd} 
\frac{1}{16} \sum_{A,B} \overline{ (\gamma_S \otimes \xi_F) }_{AB}
\frac{1}{16} \sum_{A',B'} (-1)^{C \cdot A'}
\overline{ (\gamma_{S'} \otimes \xi_{F'}) }_{A'B'} (-1)^{D \cdot B'}
(B'-A)_\mu f^\mu_{AB'} (k) 
\ I^{(a)}_{AB}
\end{eqnarray}
\end{widetext}
\subsection{Gauge invariance}
\label{subsec:gauge-inv}
As shown in Fig.~\ref{fig:2}, diagrams (c) and (d) of Fig.~\ref{fig:1}
can be expressed as a sum of diagrams (e) and (f).
As a consequence of this, diagrams (b) and (f) lead to a
gauge-invariant form of penguin operator and diagrams (a) and (e)
lead to a gauge-invariant bilinear operator.
%
%

%
First, let us derive the relationship shown in Fig.~\ref{fig:2}.
We begin with a simple identity for diagram (b):
\begin{widetext}
\begin{eqnarray}
\sum_\mu  2 \sin( \frac{k_\mu}{2} ) \cdot H^{I,(b)}_\mu (k)
&=& (ig) T^I_{cd} 
\overline{\overline{ ( \gamma_{S'} \otimes \xi_{F'} ) }}_{C'D'}
\cdot \frac{1}{16} \sum_{AB} 
\overline{ ( \gamma_S \otimes \xi_F ) }_{AB}
\cdot \Big[ \exp( i k \cdot A ) - \exp( i k \cdot B ) \Big]
\cdot I^{(a)}_{AB}
\label{eq:(b):gauge-inv}
\end{eqnarray}
\end{widetext}
The continuum correspondence of the above Eq.~(\ref{eq:(b):gauge-inv})
vanishes but the RHS of  Eq.~(\ref{eq:(b):gauge-inv}) is not zero.
Then, what is going on with the gauge invariance?
It turns out that there exists an additional term which cancels
out this term, which is the focal point of this section.
The same kind of identities for diagrams (c) and (d) can be expressed
in terms of diagrams (e) and (f).
\begin{widetext}
\begin{eqnarray}
&\sum_\mu& 2 \sin( \frac{k_\mu}{2} ) 
\cdot \Big( H^{I,(c)}_\mu (k)
+ H^{I,(d)}_\mu (k) \Big) 
= \sum_\mu 2 \sin( \frac{k_\mu}{2} ) 
\cdot \Big( H^{I,(e)}_\mu (k)
+ H^{I,(f)}_\mu (k) \Big)
\label{eq:(c)+(d)=(e)+(f)}
\end{eqnarray}
\end{widetext}
Here, the $H^{I,(e)}_\mu (k)$ and $H^{I,(f)}_\mu (k)$ are defined as
\begin{widetext}
\begin{eqnarray}
H^{I,(e)}_\mu (k) &=&  (+ig) \ T^I_{cd} \ 
\frac{1}{16} \sum_{A',B'} (-1)^{C' \cdot A'}
\overline{ (\gamma_{S'} \otimes \xi_{F'}) }_{A'B'} 
\cdot
(-1)^{D' \cdot B'}
(B'-A')_\mu f^\mu_{A'B'} (k) 
\cdot
\frac{1}{16} \sum_{A,B} \overline{ (\gamma_S \otimes \xi_F) }_{AB}
\ I^{(a)}_{AB}
\nonumber \\
\label{eq:H(e)}
\\
H^{I,(f)}_\mu (k) &=&  (+ig) \ T^I_{cd} \ 
\overline{\overline{ ( \gamma_{S'} \otimes \xi_{F'} ) }}_{C'D'}
\cdot
\frac{1}{16} \sum_{A,B} \overline{ (\gamma_S \otimes \xi_F) }_{AB}
(B-A)_\mu f^\mu_{AB} (k) 
\cdot
I^{(a)}_{AB}
\label{eq:H(f)}
\end{eqnarray}
\end{widetext}
A derivation of the identity given in Eq.~(\ref{eq:(c)+(d)=(e)+(f)})
is given in Appendix \ref{app:derive:gauge-inv}.
%

%
%
Note that $H^{I,(e)}_\mu (k)$ can be factorized into two bilinears:
one bilinear has a single gluon emitted and the other bilinear
forms a closed fermion loop identical to that of diagram (a).
Unlike the continuum operators, the staggered operators are non-local
and gauge links must be inserted between the quark and anti-quark
fields to make them gauge invariant.
Thus, at ${\cal O}(g)$, we must have $H^{I,(e)}_\mu (k)$
to keep the gauge invariance of the spectator bilinear.
Of course, this is completely a lattice artifact, which will vanish
in the limit of zero lattice spacing, $a=0$.
Let us explain this bilinear mixing in detail in the next
subsection.

\subsection{Bilinear mixing}
\label{subsec:bilinear}
Here, we combine diagrams (a) and (e) and show that they
form a gauge-invariant bilinear.
\begin{widetext}
\begin{eqnarray}
H^{I,(a)}_\mu + H^{I,(e)}_\mu(k)
&=& \frac{1}{16} \sum_{A',B'} (-1)^{C' \cdot A'}
\overline{ (\gamma_{S'} \otimes \xi_{F'}) }_{A'B'} (-1)^{D' \cdot B'}
\cdot 
\Big[ \delta_{cd} + (i g a) \ T^I_{cd} (B'-A')_\mu f^\mu_{A'B'} (k) \Big] 
\nonumber \\ & & \cdot
\frac{1}{16} \sum_{A,B} \overline{ (\gamma_S \otimes \xi_F) }_{AB}
\ I^{(a)}_{AB}
\label{eq:bilinear:gi}
\end{eqnarray}
\end{widetext}
Here, the term enclosed in square brackets is nothing but an expansion of
gauge link ${\cal V}(y_A, y_B)$ in powers of the gauge coupling $g$.
Note that the closed loop part behaves as a constant defined as
\begin{equation}
X_{(a+e)} = 
\frac{1}{16} \sum_{A,B} \overline{ (\gamma_S \otimes \xi_F) }_{AB}
\ I^{(a)}_{AB}
\end{equation}
Hence, the combined result of diagrams (a) and (e) becomes
\begin{eqnarray}
& & H^{I,(a)}_\mu + H^{I,(e)}_\mu(k) \longrightarrow 
\nonumber \\
& & \frac{ X_{(a+e)} }{a^3}   \Big[
\bar{\chi} (y_A) \overline{ (\gamma_{S'} \otimes \xi_{F'} ) }_{AB}
 {\cal V}(y_A, y_B) \chi(y_B) \Big]
\end{eqnarray}
Naturally, the next question would be what are the possible spins and
flavors for ($\gamma_S \otimes \xi_F$) in the closed loop.
Since we are talking about the vacuum diagram, it must have the
vacuum quantum number.
Hence, the natural choice would be that $(\gamma_S \otimes \xi_F) = (1
\otimes 1)$.
Another question would be how reliable is the coefficient $X_{(a+e)}$.
Since we are talking about the divergent coefficient, hence, the
perturbative determination of $X_{(a+e)}$ is highly unreliable
because the contribution from the truncated terms are also 
divergent as we approach to zero lattice spacing.
Hence, this coefficient must be determined using a non-perturbative
method.
\subsection{Penguin operator mixing}
Here, we want to present the main results of Penguin diagrams: mixing
with penguin operators at the one-loop level.
First, we address the issue of the gauge invariance.
Basically, we want to show how the RHS of
Eq.~(\ref{eq:(b):gauge-inv}) cancels out.
From the definition of $H^{I,(f)}_\mu (k)$ given in Eq.~(\ref{eq:H(f)})
it is easy to show the following Ward identity:
\begin{eqnarray}
\sum_\mu  2 \sin( \frac{k_\mu}{2} ) \cdot \Big( 
H^{I,(b)}_\mu (k) + H^{I,(f)}_\mu (k) \Big) = 0
\label{eq:H(b+f):wi}
\end{eqnarray}
This illustrates that, unlike the continuum, we need diagram (f) to
keep the gauge invariance of diagram (b), which is a pure lattice
artifact originating from using staggered fermions when constructing
operators on the lattice.
Now, we turn to explicit calculation of the mixing with penguin
operators.
First, we define $I^{(c)}_{AB}$ in terms of $I^{(a)}_{AB}$ in
Eq.~(\ref{eq:I(a):1}) as follows:
\begin{eqnarray}
I^{(c)}_{AB} &\equiv& (B-A)_\mu \cdot I^{(a)}_{AB} 
\nonumber \\
&=& \int_p \ \exp( i p \cdot (A-B) ) 
\nonumber \\ & & \cdot
\Big\{ [S_F(p)] \cdot \cos(p_\mu) 
\overline{ (\gamma_\mu \otimes 1) }
[S_F(p)] \Big\}_{BA}
\label{eq:I(a):2}
\end{eqnarray}
Using $I^{(c)}_{AB}$, we can collect diagrams (b) and (f) into such a
form that all of the nice features necessary for the gauge invariance
and the infra-red behavior are visible at a glance.
\begin{widetext}
\begin{eqnarray}
H^{I,(b+f)}_\mu(k) &\equiv& H^{I,(b)}_\mu (k) + H^{I,(f)}_\mu (k) 
\nonumber \\ &=& 
( - i g ) \ T^I_{cd} \ 
\overline{ \overline{ (\gamma_{S'} \otimes \xi_{F'}) } }_{C'D'}
\frac{1}{16} \sum_{AB} 
\cdot
\overline{ ( \gamma_S \otimes \xi_F ) }_{AB}
\exp\Big( i \frac{k}{2} \cdot (A+B) \Big) 
\cdot
\Big( I^{(b)}_{AB}(k) - I^{(c)}_{AB} \Big)
\end{eqnarray}
\end{widetext}
This result is identical to that originally presented in
\cite{ref:sharpe:2}.
Regarding the infrared behavior, note that 
\begin{equation*}
\lim_{k \rightarrow 0} I^{(b)}_{AB}(k) = I^{(c)}_{AB}
\end{equation*}
Hence, this confirms that $H^{I,(b+f)}_\mu(k)$ vanishes when $k=0$.
In addition, the quark mass behaves as an infra-red regulator
such that the integrals of $I^{(b)}_{AB}(k)$ and $I^{(c)}_{AB}$ are 
infra-red safe.
As shown in Eq.~(\ref{eq:H(b+f):wi}), $H^{I,(b+f)}_\mu(k)$ also satisfies
the Ward identity coming from gauge invariance:
\begin{equation}
\sum_\mu \sin( \frac{k_\mu}{2} ) \cdot H^{I,(b+f)}_\mu(k) = 0
\label{eq:WI:gauge inv}
\end{equation}
for all values of $k$.
Now we turn to explicit calculation of $H^{I,(b+f)}_\mu(k)$ in such a
form as we can use for matching to the continuum.
The mixing contributions comes not only from $k \sim 0$ but also, in
principle, from $k \sim \pi/a \cdot A$ for any arbitrary hypercubic vector
$A$.
Here, we first consider $k \sim 0$ as in the continuum.  
\begin{eqnarray}
H^{I,(b+f)}_\mu(k) &=& ( - ig ) T^I_{cd} \ 
\overline{ \overline{ ( \gamma_{S'} \otimes \xi_{F'} ) }}_{CD}
\nonumber \\ & & \cdot
\frac{1}{16} \sum_{A,B} 
\overline{ ( \gamma_{S} \otimes \xi_{F} ) }_{AB}
[P_\mu(k) ]_{BA}
\\ {}
[P_\mu(k) ]_{BA} &=&
- \frac{i}{ (4\pi)^2 } \sum_{\alpha,\nu,\rho}
\epsilon_{\mu\alpha\nu\rho}
k_\alpha \overline{ ( \sigma_{\nu\rho} \otimes \xi_{\nu5} ) }_{BA} 
\cdot I_a
\nonumber \\ & &
+ \frac{im}{ (4\pi)^2 } \sum_{\alpha} k_\alpha 
\overline{ ( \sigma_{\mu\alpha} \otimes 1 ) }_{BA} 
\cdot I_b
\nonumber \\ & &
+ \frac{i^2}{ (4\pi)^2 } \sum_{\alpha}
( \delta_{\mu\alpha} k^2 - k_\mu k_\alpha )
\overline{ ( \gamma_{\alpha} \otimes 1 ) }_{BA}
\cdot I_c
\nonumber \\
\label{eq:H(b+f):2}
\end{eqnarray}
where the lattice-regularized finite integrals $I_a$, $I_b$ and
$I_c$ are given in Appendix \ref{app:sec:finite-integrals}.
Note that this result for the improved staggered fermions is identical
to that for the unimproved staggered fermions presented in
\cite{ref:sharpe:2}.
This equivalence will be discussed in detail later when we present
{\bf Theorem 1}.
The first term in Eq.~(\ref{eq:H(b+f):2}) describes mixing of
the four fermion operator with a bilinear with gluon emission:
\begin{eqnarray}
& & [ \bar{\chi}_s \
 \overline{ ( \gamma_{S'} \otimes \xi_{F'} ) } \ 
\chi_d ]
[ \bar{\chi}_u \ 
\overline{ ( \sigma_{\nu\rho} \otimes \xi_{\nu5} ) } \ 
\chi_u ]
\nonumber \\
& & \longrightarrow
[\bar{\chi}_s \ 
\overline{ ( \gamma_{S'} \otimes \xi_{F'} ) } \ 
\widetilde{F}_{\nu\rho} \ 
\chi_d]
\end{eqnarray}
This corresponds to mixing with a dimension 5 operator.
From the standpoint of physics, this mixing belongs to a class of
unphysical operators because none of the operators of our interest
possesses a flavor structure of $ \xi_{\nu5} $.
Similarly, the second term in Eq.~(\ref{eq:H(b+f):2}) corresponds to
mixing of the four-fermion operators with a bilinear:
\begin{eqnarray}
& & [ \bar{\chi}_s \
 \overline{ ( \gamma_{S'} \otimes \xi_{F'} ) } \ 
\chi_d ]
[ \bar{\chi}_u \ 
\overline{ ( \sigma_{\mu\alpha} \otimes 1 ) } \ 
\chi_u ]
\nonumber \\
& & \longrightarrow
m [\bar{\chi}_s \ 
\overline{ ( \gamma_{S'} \otimes \xi_{F'} ) } \ 
F_{\mu\alpha} \ 
\chi_d]
\end{eqnarray}
This represents mixing with a dimension 6 bilinear which has 
its correspondence in the continuum.
However, this operator vanishes in the chiral limit and so
it corresponds to a higher order in the chiral perturbation.
In addition, the spin structure of the tensor does not appear in
the original set of operators of our interest and so this mixing
can occur, if any, at order $g^4$.
Hence, by convention, this term is dropped from the analysis
\cite{ref:buras:1,ref:sharpe:2}.
%

It is the third term in Eq.~(\ref{eq:H(b+f):2}) that corresponds to
mixing with penguin operators.
To complete the operator construction, the gluon needs to be connected
to the external fermion line as follows.
\begin{eqnarray}
G_{(b+f)} &=&
 - \frac{1}{N_f} \ \int_k \  H^{I,(b+f)}_\mu(k) 
\ \cdot \ D^{IJ}_{\mu\nu}(k) 
\nonumber \\ & & \cdot \  
V^J_\nu (p + \pi_C, -q-\pi_D, -k)
\label{eq:G(b+f):1}
\end{eqnarray}
where $V^J_\nu$ corresponds to the fermion vertex emitting one
gluon in \cite{ref:wlee:2}.
Here, $D^{IJ}_{\mu\nu}(k)$ represents the gluon propagator
of the thin or fat links which can be collectively written
in terms of $\hat{k}_\mu = 2 \sin(k_\mu/2)$ as
\begin{eqnarray}
D^{IJ}_{\mu\nu}(k) &=& \delta_{IJ} \  \sum_{\alpha,\beta} 
h_{\mu\alpha} (k) \ h_{\nu\beta}(k) \ 
\Bigg[ \frac{ \delta_{\alpha\beta} }
{ \hat{k}^2 }
- 
\nonumber \\ & &  ( 1 - \lambda) \cdot 
\frac{ \hat{k}_\alpha \hat{k}_\beta }
{ [ \hat{k}^2 ]^2 }
\Bigg]
\end{eqnarray}
in a general covariant gauge,
where $\hat{k}^2 = \sum_{\beta} \hat{k}_\beta^2$.
%
%
%
It turns out that only the diagonal part of $h_{\mu\alpha} (k)$
contributes mainly because the off-diagonal term of $h_{\mu\alpha}
(k)$ is proportional to $\hat{k}_\mu \hat{k}_\alpha \rightarrow k_\mu
k_\alpha$ and so the contribution from the off-diagonal term of
$h_{\mu\alpha} (k)$ vanishes due to a simple identity: $
(\delta_{\mu\alpha} k^2 - k_\mu k_\alpha) \cdot k_\mu = 0$.
\begin{equation}
(\delta_{\mu\nu} k^2 - k_\mu k_\nu ) h_{\mu\alpha}
(k) = (\delta_{\mu\nu} k^2 - k_\mu k_\nu )
\delta_{\mu\alpha} h_{\mu\mu} (k)
\end{equation}
In addition, since $h_{\mu\mu}(k) = 1 + {\cal O}(a^2k^2)$, the same
identity also guarantees that the gauge fixing term proportional to
$(1-\lambda)$ also vanishes for the leading term in the limit of $k
\rightarrow 0$, which insures gauge invariance.
In summary, we can claim that in the low momentum limit of $k\rightarrow0$,
\begin{eqnarray}
(\delta_{\mu\alpha} k^2 - k_\mu k_\alpha ) \cdot D^{IJ}_{\mu\nu}(k) &=&
\delta_{IJ} \delta_{\mu\nu}
[ h_{\mu\mu}(k) ]^2
\frac{1}{k^2} 
\nonumber \\ & &
\cdot (\delta_{\mu\alpha} k^2 - k_\mu k_\alpha ) 
\end{eqnarray}
Another important ingredient is that the $k_\mu k_\alpha$ part of
$( \delta_{\mu\alpha} k^2 - k_\mu k_\alpha )$ cannot contribute
in the on-shell limit.
Using the equations of motion for the staggered fermions, we
can prove that regardless of quark mass, 
\begin{equation}
\sum_\mu \sin(\frac{k_\nu}{2}) \cdot V^J_{\nu}(p, -q, -k) = 0
\end{equation}
Hence, in the limit of small momentum, the $\delta_{\mu\nu} k_\mu
k_\alpha$ term vanishes by the equations of motion.
In addition, in the low momentum limit,
\begin{equation}
\exp\Big(i \frac{ka}{2} \cdot (A + B) \Big) =
1 + {\cal O}(ka)
\end{equation}
and we are interested only in the leading term, since the contribution
from the remaining higher dimension operators is supposed to vanish as
we approach to the continuum ($a=0$).
Now we can simplify $G_{(b+f)}$ defined in Eq.~(\ref{eq:G(b+f):1})
at small $k$,
\begin{eqnarray}
G_{(b+f)} &=& \Big(-\frac{1}{N_f}\Big) \ 
\frac{g^2}{(4\pi)^2} \ 
\Big( \sum_I T^I_{cd} T^I_{ab} \Big) \  I_c 
\nonumber \\ & & \cdot \ 
\sum_{\mu} 
\overline{ \overline{ ( \gamma_{S'} \otimes \xi_{F'} ) }}_{C'D'}
\ \overline{ \overline{ ( \gamma_{\mu} \otimes 1 ) }}_{CD}
\nonumber \\ & & \cdot \ 
\delta_{S,\mu} \ \delta_{F,1} \ [ h_{\mu\mu}(k) ]^2
\label{eq:G(b+f):2}
\end{eqnarray}
where $k = q-p$ is strictly on shell.
From the above Eq.~(\ref{eq:G(b+f):2}), we can derive an
interesting theorem:

\bigskip
\noindent {\bf Theorem 1 (Equivalence)} \\
{\em At the one-loop level, the diagonal mixing coefficients of
penguin diagrams are identical between (a) the unimproved (naive)
staggered operators constructed using the thin links and (b) the
improved staggered operators constructed using the fat links such as
HYP (I), HYP (II), Fat7, Fat7+Lepage, and $\overline{\rm
Fat7}$}.\footnote{Note that AsqTad is NOT included in the list. In
this case, by construction the operators are made of the fat links
which are not the same as those used in the action due to the Naik
term. In addition, the choice of the fat links is open and not
unique.}

\bigskip

\noindent {\bf Proof 1.1}
In the case of the unimproved staggered operators, $h_{\mu\mu}(k) = 1$
by definition.
The improvement using the fat links such as HYP, Fat7, Lepage+Fat7,
$\overline{\rm Fat7}$, in general, leads to $h_{\mu\mu}(k)$ as defined
in Eq.~(\ref{eq:h_mat}).
The role of the additional terms proportional to $d_i$ is to suppress
the high momentum gluon interactions at the cut-off scale ($\pi/a$).
Hence, by construction, these additional terms cannot change the
dispersion in the low momentum region.
In other words, in the limit of $k\rightarrow0$,
\begin{equation}
h_{\mu\mu}(k) = 1 + {\cal O}(k^2 a^2)
\label{eq:h_11:2}
\end{equation}
Here, the ${\cal O}(k^2 a^2)$ term corresponds to higher dimension
operators, whose contribution vanishes in the limit of $a=0$.
Hence, this term is irrelevant to the penguin mixing of our interest.
It is only the leading term of Eq.~(\ref{eq:h_11:2}) that contributes
to the mixing with penguin operators.
The $[h_{\mu\mu}(k)]^2$ term is, if any, the only possible source of
difference introduced by the improvement.
However, the contribution from $[h_{\mu\mu}(k)]^2$ is, by
construction, identical in the low momentum limit before and after the
improvement.
Therefore, this leads us to the conclusion that the mixing
coefficients with penguin operators must be identical for the
staggered operators constructed using both thin links and fat links
such as HYP, Fat7, Lepage+Fat7, and $\overline{\rm Fat7}$.
This completes the proof of the theorem.
\bigskip

As a consequence of {\bf Theorem 1}, it is trivial to obtain the
diagonal mixing coefficients from Eq.~(\ref{eq:G(b+f):2}).
\begin{equation}
Z^{\rm Latt}_{ii} = - \frac{1}{N_f} I_c
\end{equation}
where $i$ represents $(\gamma_\mu \otimes 1)$.
This is our final result.

\subsection{High momentum gluons and off-diagonal mixing}
%
%
By construction, gluons carrying a momentum close to $k\sim\pi/a$ are
physical in staggered fermions and lead to taste-changing
interactions, which is a pure lattice artifact.
Let us consider a vertex where a gluon carries a momentum $k + \Pi_C$
where $\Pi_C \equiv \pi/a \cdot C$ and a quark/anti-quark has momentum $(p
+ \Pi_A)$/$(q + \Pi_B)$ respectively.
Here we assume that $|k|,|p|,|q| \ll \pi/a $.
This vertex can be expressed as follows:
\begin{eqnarray}
& & h_{\mu\nu} (k + \Pi_C) \cdot V_\nu^J(p+\Pi_A, -q - \Pi_B, k + \Pi_C) = 
\nonumber \\
& & \hspace*{5mm}
( - ig ) T^J \bar{\delta}(p-q+k) 
\overline{ \overline{ (\gamma_{\nu\tilde{C}} 
\otimes \xi_{\tilde{C}} ) } }_{A,B}
\nonumber \\ & & \hspace*{5mm} \cdot
\Big[ \frac{ 1 + (-1)^{C_\nu} }{2} + {\cal O}(ka) \Big] h_{\mu\nu}(k+\Pi_C)
\label{eq:off-diagonal}
\end{eqnarray}
Here, obviously we need to choose $C_\nu = 0$. 
In other words, the longitudinal mode is not allowed to carry high
momentum in the gluon vertex mainly because this is unphysical
and violates helicity conservation.
In the case of unimproved (naive) staggered fermions,
$h_{\mu\nu}(k+\Pi_C) = \delta_{\mu\nu}$ regardless of $\Pi_C$.
Hence, it is permissible to mix with the taste $\xi_{\tilde{C}}$ (we
call this off-diagonal mixing below) and the mixing coefficient is
substantial \cite{ref:sharpe:2}.
In contrast, in the case of the improved staggered fermions using the
fat links of our interest such as Fat7, $\overline{\rm Fat7}$ and HYP
(II),
\begin{equation}
h_{\mu\nu}(k+\Pi_C) = 0 + {\cal O}(k^2 a^2)
\end{equation}
when $C_{\rho\ne \nu} = 1$ in at least one transverse
direction.
The vertex also vanishes when $C_{\nu} = 1$.
Hence, this off-diagonal mixing is absent at the one-loop
level.
Since we adopt either $\overline{\rm Fat7}$ or HYP (II) as our
improvement scheme in our numerical study, we rejoice in this absence
of unphysical off-diagonal mixing when we analyze the data.
In the case of the improvement using HYP (I) and Fat7+Lepage,
$h_{\mu\nu}(k+\Pi_C)$ does not vanish exactly but it is
significantly suppressed.
Correspondingly, the off-diagonal mixing is similarly suppressed.
\subsection{Tadpole improvement}
In \cite{ref:wlee:1}, the procedure of tadpole improvement for
the staggered four--fermion operators is presented.
The tadpole improvement factor is given, basically, in powers of
$u_0$.
In perturbation, the contribution from this is of order $g^2$ so that
only the tadpole improvement of the original operator at the tree level
can contribute at the $g^2$ order.
Hence, the one-loop result for penguin operators are of order $g^2$
and the tadpole improvement can change the result only at the order of
$g^4$.
In other words, the tadpole improvement corrections included when we
calculate the current-current diagrams are complete at $g^2$ order and
so there is no additional correction from the tadpole improvement to
the penguin diagrams.
This argument holds valid not only for the diagonal mixing but also
for the off-diagonal mixing.

\section{Conclusion}
\label{sec:conclude}
We have studied penguin diagrams for various improved staggered
fermions at the one-loop level.
%
%
The diagonal mixing occurs only when the original operator has the
spin and taste structure of $( \gamma_\mu \otimes 1)$ regardless of
that of the spectator bilinear.
%
%
The main result summarized in {\bf Theorem 1} is that the diagonal
mixing coefficient is identical between the unimproved staggered
operators and the improved staggered operators constructed using fat
links such as Fat7, Fat7+Lepage, $\overline{\rm Fat7}$, HYP (I) and
HYP (II).
This is a direct consequence of the fact that the contribution from
the improvement changes only the mixing with higher dimension
operators and off-diagonal operators, which are unphysical.
However, Theorem 1 has such a limitation that it does not apply
directly to the case of the AsqTad staggered formulation, in which
case there is an ambiguity of choosing the fat links for the
operators.
%
%
In addition, the mixing with off-diagonal operators vanishes for Fat7,
$\overline{\rm Fat7}$ and HYP (II).
In the case of Fat7+Lepage and HYP (I), the off-diagonal mixing is
significantly suppressed by the factor of $[ h_{\mu\mu} ]^2$.
%
%

%
The results of this paper, combined with those of the current-current
diagrams in \cite{ref:wlee:5}, provide a complete set of matching for
$\epsilon'/\epsilon$ with all corrections of ${\cal O}(g^2)$ included.
In our numerical study of the CP violation, we adopt $\overline{\rm
Fat7}$ and HYP (II).
It turns out that this choice has one additional advantage of the
absence of off-diagonal mixing in penguin diagrams as well as those
advantages presented in \cite{ref:wlee:0,ref:wlee:3,ref:wlee:5}.
%


\section*{Acknowledgements}
\label{sec:acknowledge}
We would like to express our sincere gratitude to S.~Sharpe for his
consistent encouragement and helpful discussion on this paper.
We would like to thank T.~Bhattacharya, N.~Christ, G.~Fleming,
R.~Gupta, G.~Kilcup and R.~Mawhinney for their support on the
staggered $ \epsilon' / \epsilon $ project.
This work was supported in part by the BK21 program at Seoul National
University, by an Interdiciplinary Research Grant of Seoul National
University and by KOSEF through grant R01-2003-000-10229-0.
\bigskip

\appendix
\section{Derivation of Eq.~(\ref{eq:(c)+(d)=(e)+(f)})}
\label{app:derive:gauge-inv}
First, we define the common factor $Y$ as
\begin{eqnarray}
Y &=& (ig) T^I_{cd}
\frac{1}{16} \sum_{A'B'} (-1)^{C'\cdot A'}
\overline{ ( \gamma_{S'} \otimes \xi_{F'} ) }_{A'B'}
(-1)^{D'\cdot B'}
\nonumber \\ & &
\frac{1}{16} 
\sum_{AB} \overline{ ( \gamma_S \otimes \xi_F ) }_{AB}
\ I^{(a)}_{AB}
\end{eqnarray}
Using this notation, we can simplify the identities as follows.
\begin{eqnarray}
\sum_\mu  2 \sin( \frac{k_\mu}{2} ) \cdot H^{I,(c)}_\mu (k)
&=& Y \cdot \Big[ e^{i k \cdot B} - e^{ i k \cdot A' } \Big]
\\
\sum_\mu  2 \sin( \frac{k_\mu}{2} ) \cdot H^{I,(d)}_\mu (k)
&=& Y \cdot \Big[ e^{ i k \cdot B' } - e^{ i k \cdot A } \Big]
\\
\sum_\mu  2 \sin( \frac{k_\mu}{2} ) \cdot H^{I,(e)}_\mu (k)
&=& Y \cdot \Big[ e^{ i k \cdot B' } - e^{ i k \cdot A' } \Big]
\\
\sum_\mu  2 \sin( \frac{k_\mu}{2} ) \cdot H^{I,(f)}_\mu (k)
&=& Y \cdot \Big[ e^{ i k \cdot B } - e^{ i k \cdot A } \Big]
\end{eqnarray}
Using these identities, it is easy to derive
Eq.~(\ref{eq:(c)+(d)=(e)+(f)}).
\section{Finite integrals}
\label{app:sec:finite-integrals}
We use the following abbreviation to represent the integration
measure and its normalization factor.
\begin{equation}
\int_p \equiv (16\pi^2) \Pi_{\mu} \int^{\pi}_{-\pi} \frac{dp_\mu}{2\pi}
\end{equation}
Using this notation, we can express $I_a$, $I_b$ and $I_c$ as follows.
\begin{eqnarray}
I_a &=& \int_p F^2(p) \cos^2(p_\mu) \cos^2(p_\alpha) \sin^2(p_\nu)
\nonumber \\
    &=& +11.2293(3)
\\
I_b &=& \int_p F^2(p) \cos^2(p_\mu) \cos^2(p_\alpha)
\nonumber \\
    &=& 16 \Big( - \ln( 4 m^2 a^2 ) - \gamma_E + F_{0000} \Big) 
 - 40.7773(6) 
\nonumber \\ & & + {\cal O}(m^2 a^2)
\\
I_c &=& \int_p \frac{1}{6} F^2(p) 
\Big[ 2 - \sin^2(p_\mu) - \sin^2(p_\alpha)
\nonumber \\ & & 
- \sin^2(p_\mu) \sin^2(p_\alpha) \Big]
\nonumber \\
    &=& \frac{16}{3} \Big( - \ln( 4 m^2 a^2 ) - \gamma_E + F_{0000} \Big)
    - 9.5147(1) 
\nonumber \\ & & + {\cal O}(m^2 a^2)
\end{eqnarray}
where $\mu \ne \alpha \ne \nu$ and $F(p)$ is defined as
\begin{equation}
F(p) = \frac{1}{ \sum_{\mu} \sin^2(p_\mu) + (ma)^2 } \,.
\end{equation}
These integrals are also given in \cite{ref:sharpe:2} and the
results are consistent with each other.
%

\bibliography{paper}

\end{document}